\documentstyle[11pt,aaspp4,flushrt]{article} 
\input psfig
\begin{document}
\title{1/2 Law for Non-Radiative Accretion Flow} 
\author{Andrei Gruzinov}
\affil{Physics Department, New York University, 4 Washington Place, New York, NY 10003}
\affil{Institute for Advanced Study, School of Natural Sciences, Princeton, NJ 08540}

\begin{abstract}
At low densities, when radiation of an accreting gas is dynamically unimportant, quasi-spherical non-radiative accretion flow is formed around a compact object. We show that the density of the flow is $n\propto r^{-1/2}$. The gas compression is much weaker than the standard Bondi's, $n\propto r^{-3/2}$, because most of the gas does not reach the compact object. It can either circulate indefinitely long or it can be launched to infinity.  

\end{abstract}
\keywords{accretion, accretion disks -- hydrodynamics -- magnetohydrodynamics}

\section{Introduction}
Black holes in galactic centers might be so dim because the rate of mass accretion is much smaller than the Bondi rate (Blandford and Begelman 1999, Gruzinov 1998, 1999). At small gas densities, radiation is dynamically unimportant, and a quasi-spherical non-radiative accretion flow (NRAF) is formed near a compact object. A fundamental question regarding NRAFs is what is the mass accretion rate, or equivalently, how does the gas density $n$  depend on $r$, the distance from the compact object. Here we show that $n\propto r^{-1/2}$ is a logical possibility.

\section{Self-similar NRAFs}
We neglect radiation, relativistic effects, and the gas contribution to gravity. We assume that there is a point Newtonian mass (compact object) and an infinite stationary self-similar accretion flow surrounding this compact object. More precisely, the accretion flow is statistically stationary (it is most likely turbulent, eg. Hawley et al 2001 and references therein).

Self-similarity means that all velocities $v$ (including Alfven and sound)  scale as $r^{-1/2}$. The density obeys $n\propto r^{-p}$, with the exponent $p$ to be determined in what follows. 

The fluxes of mass, energy, and angular momentum scale as
\begin{equation}
\dot{M}\propto r^2nv\propto r^{3/2-p},
\end{equation}
\begin{equation}
\dot{E}\propto r^2nv^3\propto r^{1/2-p},
\end{equation}
\begin{equation}
\dot{J}\propto r^2nrv^2\propto r^{2-p}.
\end{equation}
Since the flow is stationary, all the three fluxes must be constant. The fluxes have different exponents, and they can all be constant only in the following four cases:
\begin{itemize}
\item {\bf I}.~~ $\dot{E}=0$, $\dot{J}=0$, $p=3/2$, $\dot{M}<0$
\item {\bf II}.~ $\dot{M}=0$, $\dot{J}=0$, $p=1/2$, $\dot{E}>0$,
\item {\bf III}. $\dot{M}=0$, $\dot{E}=0$, $p=2$
\item {\bf IV}.~ $\dot{M}=0$, $\dot{E}=0$, $\dot{J}=0$, $1/2<p<3/2$
\end{itemize}

Are any of these cases realizable? 

We know that {\bf I} is realizable, at least under spherical symmetry. This is the Bondi (1952) flow. Bondi flow is ideal, unmagnetized, and non-rotating. We do not know if it is realizable in nature.

{\bf II} is probably realizable. A particular case of {\bf II}  is CDAF (convection dominated accretion flow, Quataert \& Gruzinov 2000a, Narayan et al 2000, Stone et al 1999, Igumenshchev \& Abramowicz 1999). 

{\bf III} is probably not realizable because in our approximation a compact object cannot be a significant source of angular momentum. Note that at the same time the compact object can be a source of a finite energy flux (an object of a small size $\epsilon$ can produce a finite energy flux by consuming mass at a small rate $\propto \epsilon$). 

A particular case of {\bf IV} was suggested by Blandford and Begelman (1999). These authors believe that an ADIOS (advection dominated inflow outflow solution) with an unknown $p$ might exist.

Regardless of the ADIOS realizability, it should be clear that {\bf II} and {\bf IV} are separate classes. It is incorrect to think of {\bf II} as just one possible case, $p=1/2$, of a continuum of possible cases $1/2<p<3/2$. The very existence of CDAFs (solutions with $\dot{M}=0$, $\dot{J}=0$, $p=1/2$, $\dot{E}>0$) would be improbable if {\bf II} did not form a separate class.

{\bf II} is a self-similar solution of the first kind, meaning that the exponent $p$ is determined from dimensional analysis, while $\dot{E}$ can be obtained only from a detailed solution. {\bf IV} is a self-similar solution of the second kind, the exponent $p$ should be thought of as an eigenvalue, and can be obtained only together with the detailed solution. These matters are explained in a very interesting book of Barenblatt (1978).

\section{The 1/2 Law}

Consider the following thought experiment. There is a Newtonian black hole of mass $\sim 1$ (with $G=c=1$, a Newtonian black hole is an object of radius $\sim 1$ and escape velocity $\sim 1$). At the initial moment there is some gas at radius $\sim R$, the gas mass is $\sim m$, the gas energy is $\sim mR^{-1}$. A transient accretion flow will develop, and after a time $t\sim R^{3/2}$ there will be some changes of order unity. Consider predictions of {\bf II} and {\bf IV}.

According to {\bf II}, gas density near the hole is $\rho \sim m R^{-5/2}$, and the energy output of the hole, that is the power $P$,  should be $P\sim m R^{-5/2}$. The total energy ejected is $Pt\sim mR^{-1}$. This is comparable to a characteristic energy at $R$, and indeed will change the state of the gas at $R$ ``by order unity''.

According to {\bf IV}, gas density near the hole is $\rho \sim m R^{-3+p}$, and the energy output of the hole should be $P\sim m R^{-3+p}$. The total energy ejected is $Pt\sim mR^{-3/2+p}$. But if $p>1/2$, this is much greater than the characteristic energy at $R$. Only a precise cancellation of positive and negative energy fluxes can reduce the energy output, and this cancellation might be impossible in real flows.

The above argument is just a way of saying that a self-similar solution of the first kind is more natural than a self-similar solution of the second kind. This argument is certainly not rigorous. Barenblatt (1978) gives a large collection of self-similar solutions of the second kind. However, the most basic problems (diffusion, Kolmogorov turbulence, strong explosion, Bondi flow) are solved by the self-similarity of the first kind. One feels that self-similar solutions of the first kind are to be tried first.

\section{Discussion}

We predict that future simulations, and maybe even observations (Quataert \& Gruzinov 2000 b, c), might confirm the 1/2 Law for NRAFs. 

Our prediction is based on two different arguments. A rigorous argument of \S 2: 1/2 Law describes a realizable class of self-similar NRAFs, and therefore this NRAF can be found under different circumstances. A non-rigorous argument of \S 3: 1/2 Law flows are simpler than generic ADIOS.

Numerical simulation of a 3D NRAF was performed by Hawley et al (2001). As seen from the pressure map (their Fig. 1), relativistic effects limit a possible self-similar range to $r\gtrsim 30$ Schwarzschild radii. The simulation starts at $r\sim 100$, but the results between between 30 and 100, are not inconsistent with the 1/2 Law (their Fig. 2, the mass inflow=outflow$\propto r$). In order to check the 1/2 Law numerically, one needs a larger radial span. This should be possible to achieve by ignoring the small radii with their relativistic effects.

\acknowledgements

\end{document}